\documentclass[a4paper, twocolumn, superscriptaddress, floatfix, accepted=2024-06-18]{quantumarticle}
\pdfoutput=1
\usepackage[numbers, sort&compress]{natbib}
\usepackage{graphicx}
\usepackage{latexsym}
\usepackage{amssymb}
\usepackage{amsmath}
\usepackage{amsfonts}
\usepackage{upgreek}
\usepackage{subfigure}
\usepackage{bm}
\usepackage{bbold}
\usepackage{verbatim}
\usepackage[unicode=true,
 bookmarks=false,
 breaklinks=false,pdfborder={0 0 1},backref=false,colorlinks=true]
 {hyperref}
\usepackage{multirow}
\usepackage{color}
\usepackage{comment}
\usepackage{xcolor}
\usepackage{soul}
\usepackage{tikz}
\usepackage{tabularx}
\usepackage{comment}
\usepackage{enumitem}
\usepackage{flushend}
\usepackage{siunitx}

\usepackage{braket}

\begin{document}

\title{Bayesian Optimization for Robust State Preparation in Quantum Many-Body Systems}

\author{Tizian~Blatz}
\email{blatz.tizian@physik.uni-muenchen.de}
\affiliation{Department of Physics and Arnold Sommerfeld Center for Theoretical Physics, Ludwig-Maximilians-Universit\"at M\"unchen, Munich D-80333, Germany}
\affiliation{Munich Center for Quantum Science and Technology (MCQST), Munich D-80799, Germany}

\author{Joyce~Kwan}
\affiliation{Department of Physics, Harvard University, Cambridge, MA 02138, USA}

\author{Julian~Léonard}
\affiliation{Vienna Center for Quantum Science and Technology, Atominstitut, TU Wien, Vienna 1020, Austria}

\author{Annabelle~Bohrdt}
\email{annabelle.bohrdt@physik.uni-regensburg.de}
\affiliation{Munich Center for Quantum Science and Technology (MCQST), Munich D-80799, Germany}
\affiliation{Institute of Theoretical Physics, University of Regensburg, Regensburg D-93053, Germany}


\begin{abstract}
New generations of ultracold-atom experiments are continually raising the demand for efficient solutions to optimal control problems.
Here, we apply Bayesian optimization to improve a state-preparation protocol recently implemented in an ultracold-atom system to realize a two-particle fractional quantum Hall state.
Compared to manual ramp design, we demonstrate the superior performance of our optimization approach in a numerical simulation -- resulting in a protocol that is $10 \times$ faster at the same fidelity, even when taking into account experimentally realistic levels of disorder in the system.
We extensively analyze and discuss questions of robustness and the relationship between numerical simulation and experimental realization, and how to make the best use of the surrogate model trained during optimization. We find that numerical simulation can be expected to substantially reduce the number of experiments that need to be performed with even the most basic transfer learning techniques. The proposed protocol and workflow will pave the way toward the realization of more complex many-body quantum states in experiments.
\end{abstract}
\maketitle

\section{Introduction}\label{sec:Introduction}
Recent advances in quantum simulation platforms based on ultracold atoms in optical lattices~\cite{Bloch2012-QuantumSimulationsUltracold, Gross2017-QuantumSimulationsUltracold} have allowed the creation of increasingly complex quantum many-body states in larger systems, necessitating improvements in state-preparation quality and speed. Well-established approaches like adiabatic ramping~\cite{Eckstein2010-NearadiabaticParameterChangesa}, manual mapping of the quantum many-body gap~\cite{Leonard2023-RealizationFractionalQuantum}, and optimal control techniques~\cite{Doria2011-OptimalControlTechnique, Glaser2015-TrainingSchrodingerCat, Koch2016-ControllingOpenQuantum, Koch2022-QuantumOptimalControl} have allowed experimental access to exciting new phenomena, such as magnetic and topological states~\cite{Simon2011-QuantumSimulationAntiferromagnetic, vanFrank2016-OptimalControlComplex, deLeseleuc2019-ObservationSymmetryprotectedTopological, Leonard2023-RealizationFractionalQuantum}. \\
In recent years, machine learning methods, such as reinforcement learning~\cite{Niu2019-UniversalQuantumControl, Paparelle2020-DigitallyStimulatedRaman}, have emerged as a powerful tool for quantum optimal control (QOC) applications.
Bayesian optimization (BO)~\cite{Shahriari2016-TakingHumanOut, Garnett2023-BayesianOptimization}, an alternative machine-learning approach, has been used to generate efficient protocols for fast creation of high-particle number Bose-Einstein condensates~\cite{Wigley2016-FastMachinelearningOnline, Vendeiro2022-MachinelearningacceleratedBoseEinsteinCondensation}, cooling in magneto-optical traps~\cite{Tranter2018-MultiparameterOptimisationMagnetooptical}, and state preparation~\cite{Mukherjee2020-BayesianOptimalControl, Xie2022-BayesianLearningOptimal}. In these works, an online optimization loop is formed between the learner, and either an experimental setup or numerical simulation. The resulting optimized protocols have been demonstrated to compare favorably to manual ramp design~\cite{Vendeiro2022-MachinelearningacceleratedBoseEinsteinCondensation} and other optimization techniques, such as differential evolution and Nelder-Mead~\cite{Xie2022-BayesianLearningOptimal, Lagarias1998-ConvergencePropertiesNelder, Das2011-DifferentialEvolutionSurvey}. BO is well suited to these tasks featuring experimental disorder or low repetition rates, as it typically requires fewer iterations, than e.g., methods based on gradient descent, and training is robust to moderate noise. \\
In this work, we use an exact-diagonalization (ED) simulation and apply BO to prepare a few-body fractional quantum Hall (FQH) state which has recently been realized in an ultracold-atom experiment~\cite{Leonard2023-RealizationFractionalQuantum}. 
This allows us to compare the optimized result to a protocol that is manually designed and has been implemented in practice.
We extensively investigate how taking disorder in the experimental system into account changes the optimal strategy and how to engineer effective control paths in a noisy environment.
Based on this analysis, we explore how exploiting classical simulation resources can boost efficiency by reducing experimental costs.
From this, we motivate a general optimization strategy, which we believe to be applicable to any experimental system in which classical simulation is available in some capacity.
These include not only the preparation of the Laughlin state~\cite{Laughlin1983-AnomalousQuantumHall}, which acts as our benchmark, but also other current issues including challenging state preparation, such as the realization of quantum spin liquid states~\cite{Broholm2020-QuantumSpinLiquids}, and cooling in Fermi-Hubbard type systems~\cite{Mazurenko2017-ColdatomFermiHubbarda, Chiu2018-QuantumStateEngineering}. \\
In contrast to other optimal-control applications realizing a phase transition~\cite{Xie2022-BayesianLearningOptimal}, e.g., the superfluid to Mott-insulator transition~\cite{Rosi2013-FastClosedloopOptimal, vanFrank2016-OptimalControlComplex, Sorensen2019-QEngineLibraryQuantum, Mukherjee2020-BayesianOptimalControl}, the FQH problem features a topologically non-trivial target state prepared by tuning a larger number of control fields (4 vs 1-2) to be optimized over time.
This more demanding problem can shed light on the prospects of machine-learning-optimized state preparation, entering a regime where simple linear ramping is impossible and an optimized protocol's robustness and generality become primary concerns.
To determine whether BO is a tool fit to tackle these challenges, our work proceeds as follows:
We discuss our optimization strategy and investigate the experimental setting motivating our work by comparing experimental data to ED simulation.
In the core part of our work, we apply BO to the simulated system and thoroughly investigate the performance of optimized ramping protocols regarding fidelity, speed, and robustness to disorder.
We conclude with a proposal that we believe will facilitate the creation of more complex states in the system under investigation and also act as a general tool to boost efficiency in QOC applications going beyond this specific system. \\

\section{Optimization Strategy}\label{sec:Optimization Strategy}
\begin{figure}[t!!]
\centering
\includegraphics[width=\linewidth]{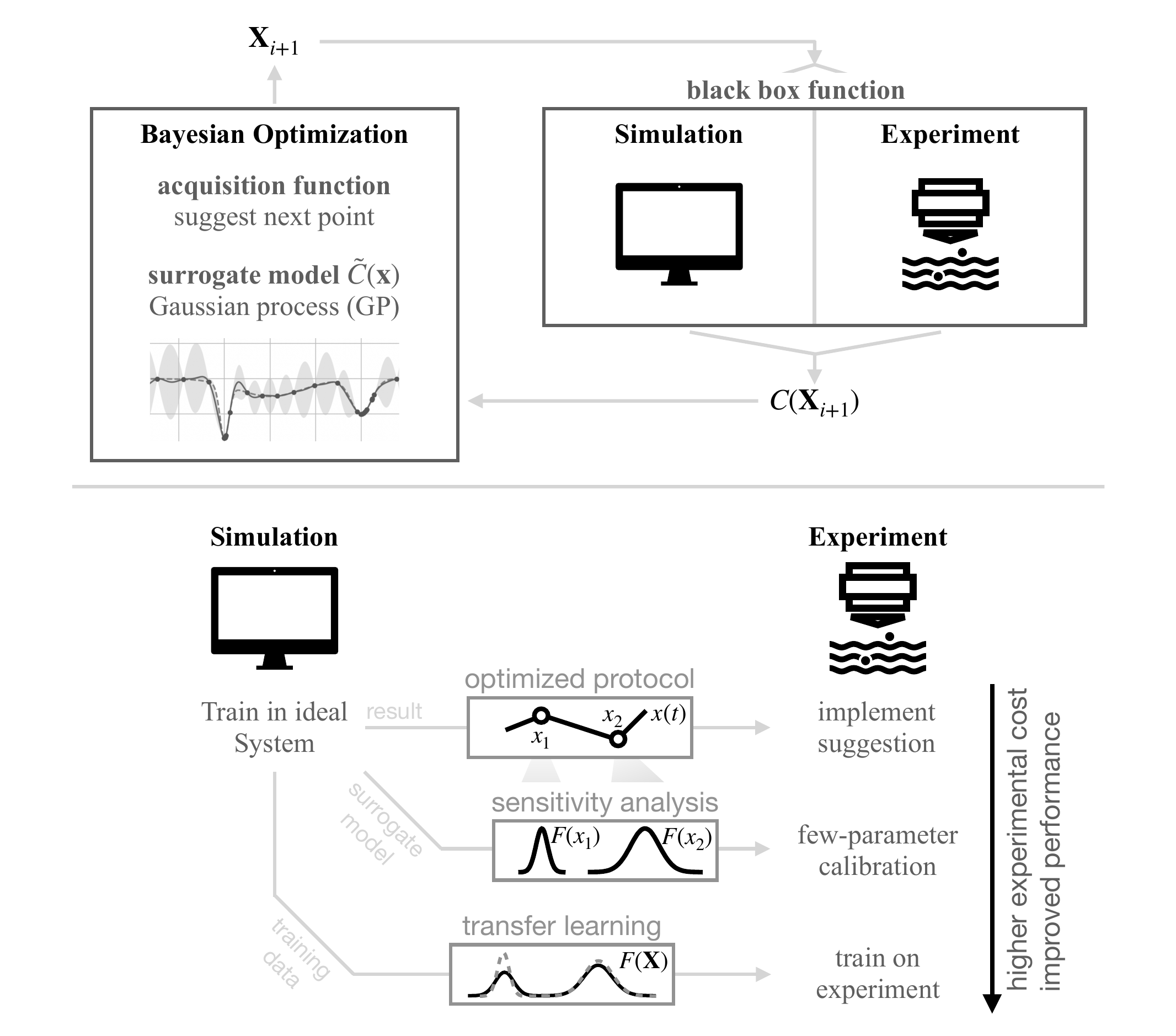}
\caption{{Optimization Strategy:}~\textbf{(top)}~Illustration of the Bayesian optimization training loop. The experiment or numerical simulation acts as a black-box function, providing a quality measure for a given ramping protocol. Based on evaluations of this function, the BO learner builds a surrogate model of the cost landscape and suggests the next protocol to implement.
\textbf{(bottom)}~Proposed optimization strategy exploiting numerical resources to boost optimization efficiency in the experiment.
Depending on the task at hand, several avenues with varying experimental costs may be explored.
The most straightforward way is to directly implement the optimization result obtained from simulation in the experiment, requiring only a single evaluation of the protocol's quality.
On top of this, the BO surrogate model can provide a sensitivity analysis that can be exploited to manually calibrate only the most sensitive protocol parameters in $\sim 10$ experimental evaluations.
Finally, when substantial adaptation to the experimental setting is necessary, online optimization can be performed in $\sim 100$ to $\sim 1000$ iterations directly with the experiment. Transfer-learning approaches can leverage numerical resources to accelerate training significantly, e.g., by initializing the surrogate model used for training in the experiment with training data from the simulation.
The prospective reduction of the number of cost-function evaluations in the experiment is crucial in the context of optical lattice systems where typical snapshot-based detection schemes require many experimental runs to determine an accurate figure of merit.
In this work, we highlight and motivate these three scenarios by comparing the numerical simulation of an idealized system to a more realistic description of the experimental setting that takes experimental disorder into account. }
\label{fig_bayesian_optimization}
\end{figure}

In this work, we implement a BO training loop as illustrated in Fig.~\ref{fig_bayesian_optimization}.
Gaussian processes (GP)~\cite{Rasmussen2005-GaussianProcessesMachine} are used as a surrogate model of the cost function or figure of merit of ramping protocols over the entire parameter space. For details about BO and our implementation, the reader is referred to Appendix~\ref{subsec:BO}. \\
Fig.~\ref{fig_bayesian_optimization} also shows the proposed optimization strategy to be used in experiments. The strategy is focused on making use of the trained surrogate model and numerical simulation to reduce experimental costs. We develop this strategy in the following sections. \\

\section{Experimental Setting}\label{sec:Experimental Setting}
Our application of BO is focused on the state-preparation problem of a bosonic two-particle FQH state recently realized in experiment~\cite{Leonard2023-RealizationFractionalQuantum}. The experiment features a ramp protocol, adiabatically\footnote{In the context of optimization, we consider an adiabatic protocol to be any strategy aiming to minimize losses of the target-state population by limiting the ramp speed to the maximum set by the inverse of the many-body gap.} transforming an initial state of two pinned particles into the delocalized FQH state.
\begin{figure}[t!!]
\centering
\includegraphics[width=\linewidth]{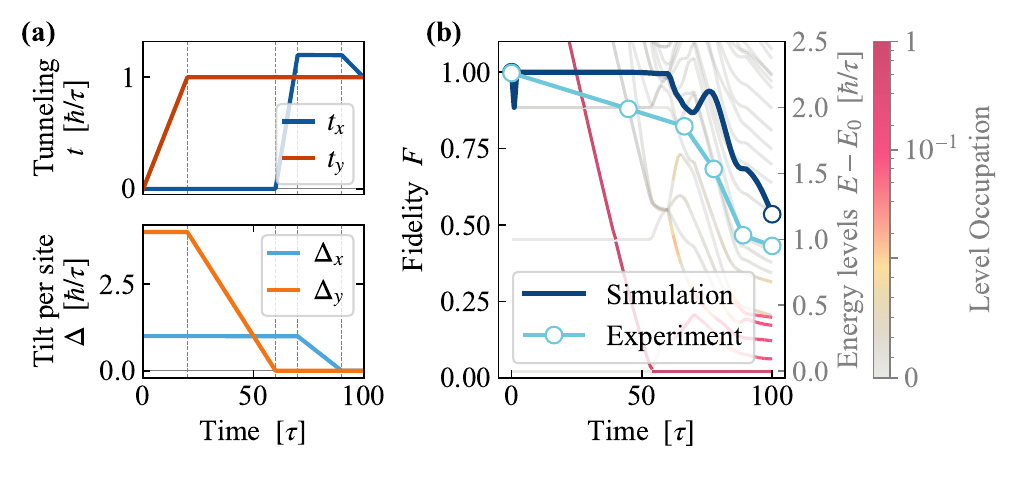}
\caption{{Experimental Ramp:}~\textbf{(a)}~Experimental ramping protocol for the $4$ lattice parameters adapted from \cite{Leonard2023-RealizationFractionalQuantum}.
\textbf{(b)}~Fidelity w.r.t. the instantaneous ground state throughout the ramp. The ED simulation is compared to experimental data~\cite{Leonard2023-RealizationFractionalQuantum}. The color-coded lines in the background show the energy levels $E - E_0$ of $\hat{H}(t)$ and their respective occupation during the simulated ramping protocol. Here, $E_0$ is the instantaneous ground-state energy.
Before $t_x$ is turned on at $T = 60 \: \tau$, the system is effectively one-dimensional, since all particles are initialized in the 1st column of the $4 \times 4$ lattice and there is no coupling in x-direction. In this regime, we define the fidelity wrt. this 1D system to avoid spurious drops to $F \sim 0$ when $\Delta_x < \Delta_y$ caused by the initial-state geometry.
}
\label{fig_experimental_ramp}
\end{figure}
\begin{figure*}[t!!]
\centering
\includegraphics[width=\textwidth]{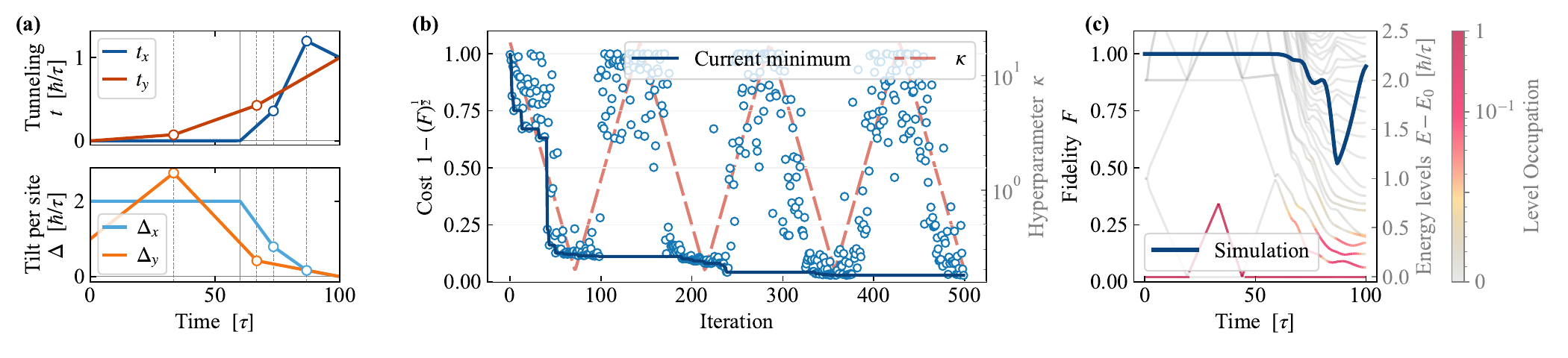}
\caption{{Optimized Ramp:}~\textbf{(a)}~Optimized ramp protocol for the $4$ lattice parameters. White dots show the parameter values controlled by the optimizer at fixed points in time (vertical dashed lines). \textbf{(b)}~Evaluated cost function at each iteration of the BO training. The visible oscillations in the evaluated fidelity follow sweeps of the BO hyperparameter $\kappa$, which tunes the strategy between exploration and exploitation (see Appendix~\ref{subsec:BO} for optimization details).
\textbf{(c)}~Fidelity w.r.t. the instantaneous ground state throughout the ramp.
The color-coded lines in the background show the energy levels of $\hat{H}(t)$ and their respective occupation during the simulated ramping protocol. All energies are w.r.t. the ground state energy which sets $E = 0$.
Before $t_x$ is turned on at $T = 60 \: \tau$, the system is effectively one-dimensional, since all particles are initialized in the 1st column of the $4 \times 4$ lattice and there is no coupling in x-direction. In this regime, we define the fidelity wrt. this 1D system to avoid spurious drops to $F \sim 0$ when $\Delta_x < \Delta_y$ caused by the initial-state geometry.
}
\label{fig_optimal_ramp}
\end{figure*}
To achieve this, four global lattice parameters in the interacting Harper-Hofstadter model~\cite{Tai2017-MicroscopyInteractingHarper, Harper1955-GeneralMotionConduction, Hofstadter1976-EnergyLevelsWave}
\begin{align}
\begin{split}
    \hat{H} = \
    &- t_x \sum_{x, y} (e^{i \phi y} \hat{a}^\dagger_{x, y} \hat{a}^{\vphantom{\dagger}}_{x+1, y} + \mathrm{H.c.})
    + \Delta_x \sum_{x, y} x \, \hat{n}_{x, y} \\
    &- t_y \sum_{x, y} (\hat{a}^\dagger_{x, y} \hat{a}^{\vphantom{\dagger}}_{x, y+1} + \mathrm{H.c.})
    + \Delta_y \sum_{x, y} y \,\hat{n}_{x, y} \\
    &+ \frac{U}{2} \sum_{x, y} \hat{n}_{x, y} (\hat{n}_{x, y} - 1)
\end{split}
\end{align}
are varied independently over time: (i, ii) the tunneling strengths $t_{x,y}$ in x- and y-direction and (iii, iv) the chemical potential gradients $\Delta_{x,y}$. Here, $\hat{a}^{(\dagger)}_{x, y}$ is the bosonic annihilation (creation) operator at site $(x, y)$ and $\hat{n}^{\vphantom{\dagger}}_{x, y} = \hat{a}^\dagger_{x, y} \hat{a}^{\vphantom{\dagger}}_{x, y}$ is the particle number operator.
The tunneling phase $\phi / 2 \pi = 0.27$ is constant throughout this work, corresponding to the Laughlin-like state~\cite{Laughlin1983-AnomalousQuantumHall} at the fractional filling $\nu = 1/2$, where $\nu$ is the filling factor.
This target state is realized in the flat lattice $\Delta_x = \Delta_y = 0$, at equal tunneling amplitudes $t_x = t_y = \hbar / \tau$.
We fix the value of the tunneling amplitudes in the target state as our unit of energy while the tunneling time $\tau$ acts as the unit of time.
In the experiment, $\tau = 4.3(1) \: \mathrm{ms}$~\cite{Leonard2023-RealizationFractionalQuantum}.
Finally, the on-site repulsion is fixed to $U = 8 \: \hbar / \tau$. \\
The ramping protocol used in the experiment was devised by mapping out the many-body gap over the four-dimensional parameter space and manually identifying a path that avoids a gap closing. In addition, the tunneling $t_x$ in x-direction is only turned on in the latter half of the experimental protocol to reduce the effects of Floquet heating~\cite{Strater2016-InterbandHeatingProcesses} resulting from the synthetic magnetic field engineered in the experiment.
The protocol is illustrated in Fig.~\ref{fig_experimental_ramp}. It serves as the primary benchmark in this work -- representing the capabilities of sophisticated adiabatic ramp design. \\
The quality of a ramp is quantified by the fidelity
\begin{align}
    F = | \langle \psi_\mathrm{final} | \psi_0 \rangle |^2 \ ,
\end{align}
where $\ket{\psi_\mathrm{final}}$ is the state after the preparation protocol, and $\ket{\psi_0}$ is the ground state of $\hat{H}$ at the target parameters $t_x = t_y = \hbar / \tau$ and $\Delta_x = \Delta_y = 0$.
This is a good figure of merit since it is experimentally accessible by inverting the ramp and measuring the fraction that has returned to the initial state using snapshots in the number basis. For the experimental protocol it was measured to be $F = 43(6)\%$~\cite{Leonard2023-RealizationFractionalQuantum}.
In Fig.~\ref{fig_experimental_ramp}, we compare the experimental data to an ED time-evolution simulation.
Even though the simulation is performed at zero temperature, does not take into account any lattice defects, and does not model the Floquet sequence employed in the experiment -- instead simulating the unperturbed effective Hamiltonian -- the fidelity-loss over time (see Fig.~\ref{fig_experimental_ramp}b)) shows qualitative agreement with the experiment.
The fidelity of $F = 53.5 \%$ observed in the simulation is not much higher than the value reported in the experiment. This supports non-adiabatic losses to excited states during the state-preparation protocol as the leading source of fidelity loss, providing a promising foundation for ramp optimization. \\

\section{Optimization}\label{sec:Optimization}
For each lattice parameter, we parametrize the time-dependent ramp by $N_s$ linear segments of equal length.
Fixing the initial and target parameters to be the same as in the experiment leads to a BO problem with $4 \times (N_s - 1)$ parameters.
Motivated by the experimental conditions and concerns about Floquet heating, we limit the search space to
\begin{align}
    0 \leq t_{x/y} \leq 1.2 \: \hbar / \tau \ \mathrm{and} \ -4 \: \hbar / \tau \leq \Delta_{x/y} \leq 4 \: \hbar / \tau \ ,
\end{align}
which corresponds to a Floquet regime that minimizes excitations to higher bands and achieves efficient dynamics. Specifically, the range for $t_{x/y}$ is chosen to limit the broadening of heating channels~\cite{Strater2016-InterbandHeatingProcesses}, as its strength is determined by the driving amplitude. We start ramping $t_x$ and $\Delta_x$ only after passing the time $T^{(0)}_x = 0.6 \: T$ where $T$ is the total ramp time.
In summary, the optimizer may choose the values of $t_y$ and $\Delta_y$ at times $T^{(1)}_y = T / 3$ and $T^{(2)}_y = 2 \: T / 3$, and the values of $t_x$ and $\Delta_x$ at times $T^{(1)}_x = T^{(0)}_x + (T - T^{(0)}_x) / 3$ and $T^{(2)}_x = T^{(0)}_x + 2 \: (T - T^{(0)}_x) / 3$. \\
After training with $500$ iterations, BO identifies an optimal ramp with $F = 94.5 \%$ as illustrated in Fig.~\ref{fig_optimal_ramp}.
Interestingly, the protocol is qualitatively similar to the one used in the experiment.
As depicted in Fig.~\ref{fig_optimal_ramp}, it achieves near-perfect fidelity due to significant gains in the ground-state population (w.r.t. the instantaneous ground state) near the final protocol time.
Such a resurgence of fidelity has also been reported in other simulation works on QOC applications~\cite{Mukherjee2020-PreparationOrderedStates}.
While captured by optimal control theory, it is outside the scope of manual design principles, such as adiabatic ramp design, which aim to minimize fidelity losses, but do not take processes that can return population to the ground state into account.
We now turn to whether this feature, observable in simulations, can be expected to be robust in experimental conditions. \\

\section{Training with Noise}\label{sec:Training with Noise}
\begin{figure}[t!!]
\centering
\includegraphics[width=\linewidth]{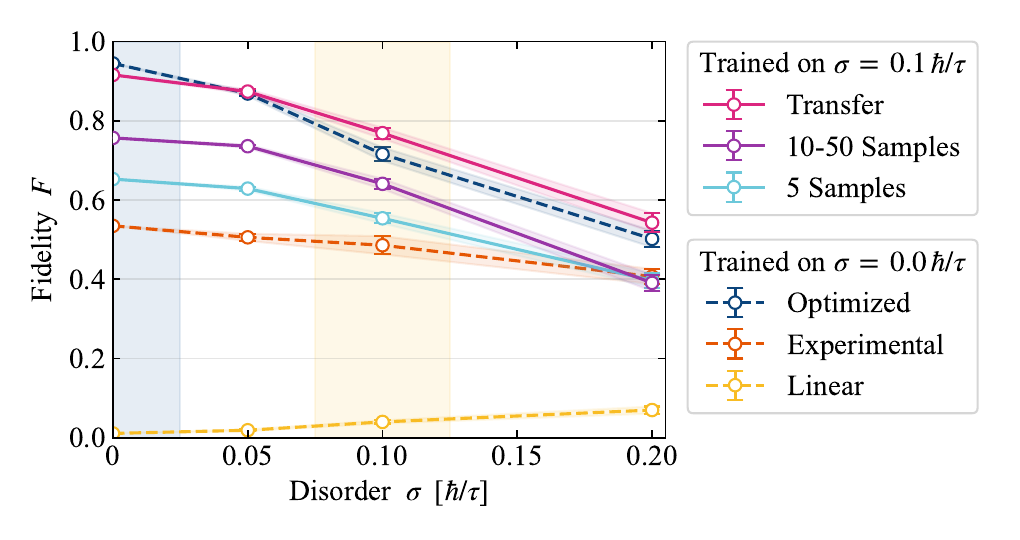}
\caption{{Experimental Noise:}~Expected fidelity $F$ of different ramping protocols at varying standard deviation $\sigma$ of the experimental disorder. The shaded areas show the low-disorder (blue) and the experimentally realistic (orange) regime.
Solid lines show ramps trained in the presence of disorder,
dashed lines show ramps designed in the ideal system.
The error bars indicate the standard error of the mean fidelity obtained from averaging over $100$ disorder realizations.
The highest fidelity at $\sigma = 0$ is achieved by optimization in the clean system.
The highest fidelity at $\sigma \geq 0.1 \: \hbar / \tau$ is achieved by training in the disordered system, initialized from training in the ideal system (`Transfer').
}
\label{fig_fidelity_noisy_ramps}
\end{figure}

To characterize robustness, and estimate the real-world fidelity we would expect from optimized protocols, we now model experimental disorder in our simulation by the sum of random and harmonic chemical potential offsets (see Appendix~\ref{subsec:Noise}).
The disorder $\mu$ is taken to be Gaussian-distributed with mean $0$ and standard deviation $\sigma = 0.1 \: \hbar / \tau$ which is an experimentally realistic disorder level.
For each numerical evaluation of a ramp's fidelity, we perform time evolution with a newly sampled realization of the disorder. \\
Optimization is now performed in the presence of this disorder to represent training in an experimental setting. To mitigate fluctuations, we take the average fidelity of $5$ disorder realizations as the cost function, which is a typical value used in the literature~\cite{Vendeiro2022-MachinelearningacceleratedBoseEinsteinCondensation, Xie2022-BayesianLearningOptimal}. After training for $1000$ iterations, we compare the optimized protocol to the manually designed ramp and the optimum found from training without disorder in Fig.~\ref{fig_fidelity_noisy_ramps}.
The comparison is performed over a range of disorder strengths, covering both the clean as well as the experimentally realistic case.
Firstly, this analysis highlights the competitiveness of the manually designed protocol and the complexity of the state preparation problem at hand.
The experimentally implemented ramp is remarkably robust and evaluates to $F = 49(2) \%$ at $\sigma = 0.1 \: \hbar / \tau$, even closer to the experimentally measured ramp fidelity and more than $10 \times$ the fidelity expected from simple linear ramping. \\
The fidelity of the ramp optimized in noisy training beats the fidelity of the manually designed protocol at all investigated disorder levels. However, it is inferior to the optimum found in training without disorder.
We conclude that the noisy evaluation of the cost function makes training significantly more challenging. \\
We confirm this by significantly increasing the number of disorder realizations averaged over for an evaluation of the cost function to an adaptive number between $10$ and $50$ -- taking more samples when the fidelity is high (see Appendix~\ref{subsec:Sampling} for more details). Doing so boosts BO performance but still does not reach the fidelity achieved in training without disorder. \\
To reach optimum performance in the noisy environment, we turn to our knowledge of the ideal system.
Usually, the surrogate model in BO is initialized by picking $\mathcal{O}(10)$ points in parameter space at random and evaluating the fidelity at these points.
As a rudimentary implementation of transfer learning, we replace this random selection with a subset of the points visited during training in the ideal system.
The most relevant points for training are chosen by a simple heuristic algorithm according to their fidelity (which is known for the clean system) and Euclidean distance in parameter space.
For details about the algorithm, see Appendix~\ref{subsec:Algorithm}.
The cost function at the chosen points is subsequently reevaluated in the presence of disorder to initialize the surrogate model. \\
With this most basic transfer-learning approach, we reach the highest fidelity achieved in the noisy environment with a fraction of the necessary number of runs in the disordered system.
Comparing this protocol with the optimized result from the clean system tells us something about the changes in the cost landscape when disorder is included.
The new optimum in the disordered system is very close in parameter space to that in the clean system, so we find disorder to only lead to a slight shift in the ideal strategy. This scenario is distinct from one realized in a previous study~\cite{Mukherjee2020-PreparationOrderedStates} 
where the optimum location drastically changed with the onset of disorder.
For a more detailed discussion of these two scenarios, see Appendix~\ref{subsec:Brute-Force} which confirms our findings as a property of the system in question -- not an artifact of optimization -- via a brute-force search and examines a special case in our system that mirrors the scenario in the literature. \\

\section{Short Ramps}\label{sec:Short Ramps}
Thus far, we have aimed to produce ramping protocols that achieve high fidelity both in an ideal as well as a noisy setting. The total ramp time $T$ has been kept fixed to that of the adiabatic experimental protocol.
However, a major advantage of QOC techniques is that they can significantly reduce the total protocol time while retaining high fidelities~\cite{Wigley2016-FastMachinelearningOnline, vanFrank2016-OptimalControlComplex}.
To leverage this advantage, we run optimization with reduced final times $T = 10 \: \tau$, and $T = 20 \: \tau$, i.e., $10 \times$, or $5 \times$ shorter than the protocols previously considered.
Based on our findings in the previous sections, we expect the global optimum of the optimization landscape to move only a little when introducing experimental disorder, so we chose to work in the ideal system. \\
Running the optimization for $500$ iterations, we discover ramping protocols with $F= 53\%$, and $F = 78\%$, respectively.
\begin{figure}[t!!]
\centering
\includegraphics[width=\linewidth]{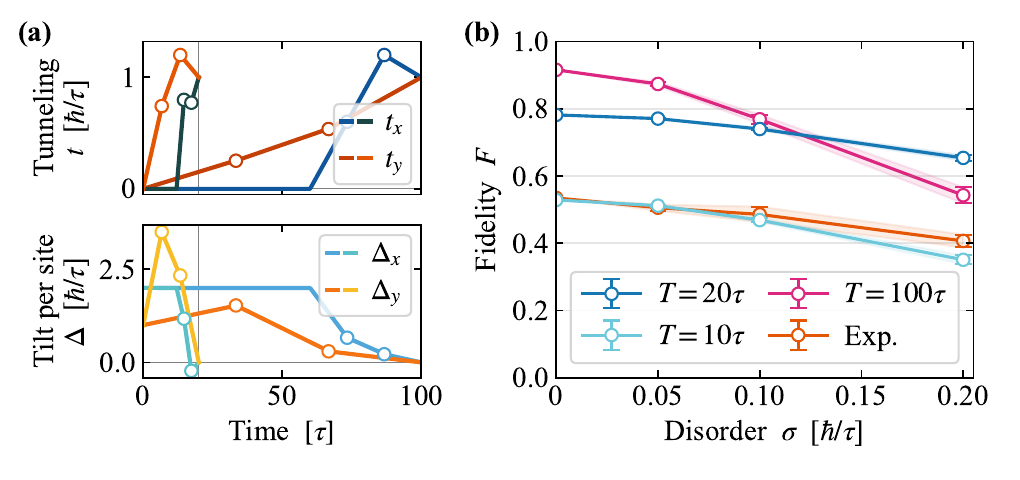}
\caption{{Short Ramps:}~\textbf{(a)}~Ramp protocols for the optimized ramps with final times $T = 20 \: \tau$ and $T = 100 \: \tau$. The ramp shown for the final time $T = 100 \: \tau$ is the most robust one considered in the previous section, i.e., the `Transfer' ramp from Fig.~\ref{fig_fidelity_noisy_ramps}. White dots show the parameter values controlled by the optimizer at fixed points in time. \textbf{(b)}~Fidelities of the ramps optimized for various final times, as well as the experimental ramping protocol (which has final time $T = 100 \: \tau$) at varying disorder level $\sigma$.
The error bars indicate the standard error of the mean fidelity obtained from averaging over $100$ disorder realizations. }
\label{fig_fidelity_short_ramps}
\end{figure}
The shorter protocols are compared to the ones with $T = 100 \: \tau$ in Fig.~\ref{fig_fidelity_short_ramps}. We find the shortest ramp to be remarkably close to the experimental protocol in terms of the resulting fidelity -- both in the clean as well as the disordered system. That is, at this fixed level of fidelity, optimization reaches a speedup by a factor of $10$. \\
We consider the ramp resulting from the ramping time $T = 20 \: \tau$ to be optimal for an experimental implementation. While the fidelity in the disorderless scenario is still significantly lower than for optimized ramps with longer ramp times, we find the protocol to be extremely robust under the introduction of disorder.
In fact, the fidelity at the strongest considered level of disorder $\sigma = 0.2 \: \hbar / \tau$ is significantly higher than for any other protocol considered in this work. Similar to previous studies \cite{vanFrank2016-OptimalControlComplex, Mukherjee2020-PreparationOrderedStates}, we attribute this result to the shorter protocol time presenting less of an opportunity for the accumulation of errors. This interpretation is supported by the significantly lower standard deviation of the fidelity in the noisy setting.
At the experimental disorder level $\sigma = 0.1 \: \hbar / \tau$, while the mean fidelity of the $T = 20 \: \tau$ ramp is similar to that of the ramp optimized for $T = 100 \: \tau$, the standard deviation is lower by more than a factor of $4$. \\

\section{Experimental Proposal}\label{sec:Experimental Proposal}
\begin{figure}[t!!]
\centering
\includegraphics[width=\linewidth]{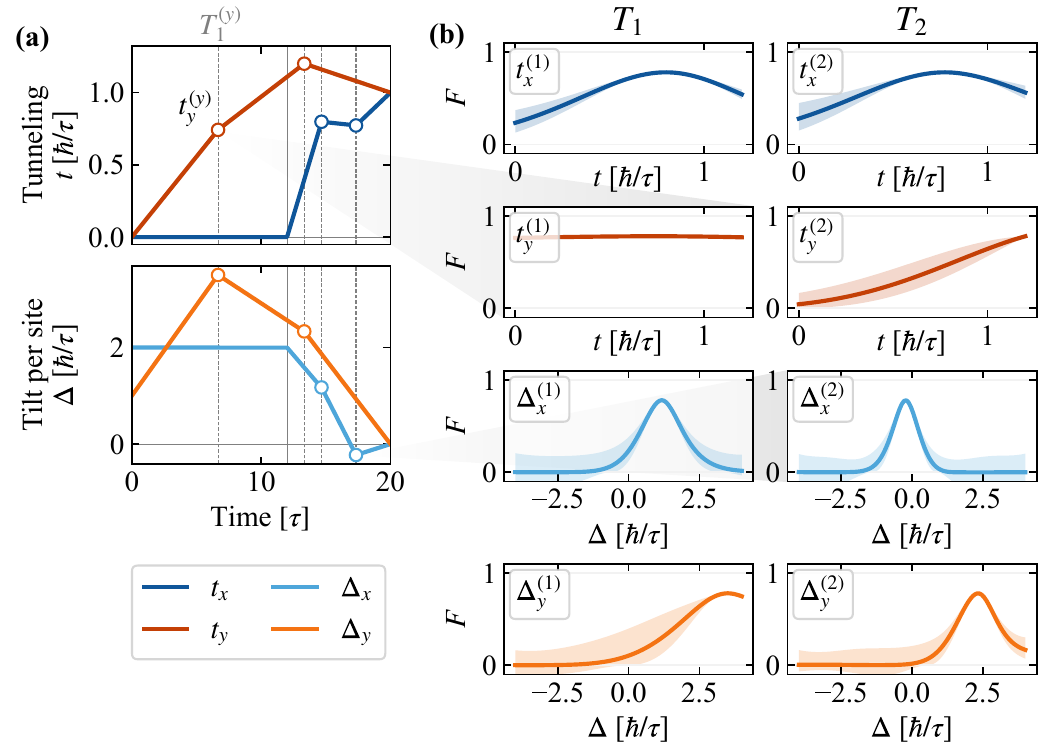}
\caption{{GP Model:}~Gaussian-process model of the cost landscape around the global optimum trained during optimization of the $T = 20 \: \tau$ ramp. \textbf{(a)}~The optimum ramping protocol identified by optimization -- characterized by $4$ control fields at $2$ time points (white dots). \textbf{(b)}~Single-parameter variations around the global optimum of the modeled cost landscape with the remaining parameters fixed to their optimum value.
The shaded regions depict the model uncertainty (one standard deviation), which is generally the lowest around the global optimum.
The sensitivity of the model to changes in different parameters varies considerably. Grey, shaded regions link the least (most) sensitive protocol parameter $t_y^{(1)}$ ($\Delta_x^{(2)}$) to the corresponding modeled optimum shape. }
\label{fig_gp_model}
\end{figure}

Based on the findings of the previous sections, we propose the following strategy and best practices for the realization of the BO approach in an experiment: \\
(i) Implement suggestion: Due to the advantages in terms of speed and robustness of shorter ramping protocols, we suggest the $T = 20 \: \tau$ ramp from the previous section for an experimental realization. Even though the training was performed on the ideal system, we believe the scheme to be directly implementable, as we found the global optimum location to be relatively insensitive to disorder. \\
Direct implementation of a numerically optimized (local adiabatic) control path has previously been used successfully to realize few-particle crystallization in an experiment~\cite{Schauss2015-CrystallizationIsingQuantum}. \\
(ii) Sensitivity analysis: Going beyond this innate robustness, we argue that a sensitivity analysis based on the BO surrogate model can reduce manual calibration in the experiment to a few parameters.
Since BO takes an increased number of samples around the discovered optimum, the model is very accurate in this region and can faithfully predict the optimum's shape in parameter space.
Thus, the surrogate model contains accurate information about the optimum's sensitivity to the different parameters.
When the optimum is only slightly shifted in parameter space by the onset of disorder, good performance can be recovered at moderate experimental cost by manually tuning only the most sensitive parameters. \\
The sensitivity analysis drawn from training the proposed $T = 20 \: \tau$ ramp is depicted in Fig.~\ref{fig_gp_model}.
By varying single parameters around the optimum, we identify the fidelity to be much more sensitive to tuning the global tilts than to changes in hopping amplitudes.
In particular, the GP model of the cost function is almost completely insensitive to the value $t_y^{(1)}$ at the first time point $T = 20/3 \: \tau$ but sharply peaked around the optimum value of $\Delta_x^{(2)}$ at the second fixed time point.
By taking advantage of the trained learner in this way, manual calibration in the experiment can be reduced from full functional dependencies to only a few parameters. \\
(iii) Transfer learning: In case sensitivity analysis fails to recover the optimum, or the system is intractable numerically, it becomes favorable to train on the experiment directly.
In Section~\ref{sec:Training with Noise}, we have demonstrated that making use of simulation data in a simple transfer-learning approach can significantly boost performance while reducing experimental costs (represented by the number of simulation runs in the disordered system).
In Section~\ref{subsec:Pfaffian} we also find that finding high-fidelity state-preparation protocols for a $4$-particle Pfaffian-like state~\cite{Palm2021-BosonicPfaffianState} on a larger $L_x = L_y = 6$ lattice is made much easier by transferring the results for the smaller state obtained in Section~\ref{sec:Optimization}. This presents an avenue that could extend efficiency gains to experimental systems beyond the reach of classical simulation.
\\
\\

\section{Discussion and Outlook}\label{sec:Discussion and Outlook}
We applied a Bayesian optimization approach to optimize parameter ramps for state preparation. The optimization scheme was demonstrated to be efficient at finding protocols that outperform manual design not only in an ideal system but also in the presence of significant disorder -- identifying a ramp that is both $5 \times$ faster as well as twice as close to ideal fidelity as the manual reference.
The superior performance holds when the training data itself is noisy, e.g., when training directly on an experiment, but at an increased cost of training. \\
We propose several approaches to decrease experimental costs in cases where numerical simulation of the system is available.
Namely, we demonstrated that optimized solutions for the ideal system can be intrinsically robust due to a significant reduction in protocol duration and the surrogate model of the fidelity-landscape naturally provides information about the most sensitive parameters to tune in an experiment.
The reductions in experimental cost achievable by leveraging numerical resources are especially valuable to systems of ultracold atoms in optical lattices, where estimating the figure of merit usually requires many experimental snapshots.
\\
Even when learning directly on the disordered system representing an experiment, prior training in a simulated environment was demonstrated to significantly reduce the total cost in Section~\ref{sec:Training with Noise}. \\
The central challenge ahead lies in extending efficient optimization to system sizes that are not tractable using numerical simulation.
Promising recent results in this regard were obtained by Y.-J. Xie et al.~\cite{Xie2022-BayesianLearningOptimal}, who report efficient transfer of BO to iteratively larger system sizes in a Heisenberg chain.
For the FQH system discussed here, we demonstrate good transferability to a larger $4$-particle state with different model parameters in Section~\ref{subsec:Pfaffian}.
In particular, this marks an example, where the substantial speedup offered by protocols optimized using machine learning can allow the preparation of a more complex quantum state within the already available coherence times of contemporary ultracold-atom experiments. \\
By using either such a transfer-learning approach or alternative schemes merging several FQH plaquettes~\cite{Palm2024-GrowingExtendedLaughlin}, we expect the optimized protocols developed here to be instrumental in realizing larger FQH states, such as the Pfaffian state, and different filling factors.
The more general optimization strategy, we believe to be widely applicable to boost experimental efficiency and improve the interface between classical and quantum simulation.


\textbf{Data and Code Availability --}
The data presented in the figures is available at \href{https://github.com/TizianBlatz/bo-state-preparation}{https://github.com/TizianBlatz/bo-state-preparation}, which also contains a minimal working example of the optimization routine.

\textbf{Acknowledgments --}
We are very grateful to F. Grusdt, F. Palm, and C. Weitenberg for fruitful discussions, and we wish to thank M. Greiner, B. Bakkali-Hassani, P. Segura, and Y. Li for valuable insights regarding the experimental system.
This research was funded by the Deutsche Forschungsgemeinschaft (DFG, German Research Foundation) under Germany’s Excellence Strategy—EXC-2111—390814868.
J.L. acknowledges support by the FWF grant no. Y-1436 and the FFG grant no. F0999896041.

\appendix



\section{Bayesian Optimization}\label{subsec:BO}
Bayesian optimization (BO)~\cite{Shahriari2016-TakingHumanOut, Garnett2023-BayesianOptimization} is an optimization method designed to find the optimum of a -- possibly noisy and expensive to evaluate -- black box function.
In our case, the task is finding the optimal ramping protocol $\hat{H}(t)$ for the time-dependent Hamiltonian that transforms a given initial state $\ket{\psi_\mathrm{initial}}$ into the target state $\ket{\psi_\mathrm{target}}$.
The quality of a given ramping protocol is quantified by the fidelity
\begin{align}
    F = | \braket{\psi(T) | \psi_\mathrm{target}} |^2 \ ,
\end{align}
where $\ket{\psi(T)}$ is the initial state evolved according to $\hat{H}(t)$ up to the final ramp time $T$. \\
To phrase this as a standard optimization problem, we consider Hamiltonians $\hat{H}(t)$ characterized by a set of parameters $\{ h_i(t) \}_i$. Then, we take each Hamiltonian parameter $h_i(t)$ as a piece-wise linear function characterized by a set of time-value tuples $\{ x_j = (t_j, v_j) \}_j$. The optimization task then becomes minimizing the cost function
\begin{align}
    C(\{ x_j \}_j) = 1 - \sqrt{F} \ .
\end{align}
To find optimal values for the parameter set $\mathbf{x} = \{ x_j \}_j$, a BO loop, based on the open-source \texttt{scikit-optimize} package, is run in the following way: First, the search is initialized by randomly choosing some $\mathbf{x}_i$ -- typically $\mathcal{O}(10)$ points -- and evaluating the corresponding cost function values $C(\mathbf{x}_i)$ -- either by running an experiment or, in our case, via numerical simulation.
We then use Gaussian processes (GP)~\cite{Rasmussen2005-GaussianProcessesMachine} to build a surrogate model $\Tilde{C}(\mathbf{x})$ of the cost function over the entire parameter space, based on the set of evaluated points.
A GP is a generalization of the Gaussian distribution, defining a distribution over functions by replacing the mean vector and covariance matrix by the mean function $m(\mathbf{x})$, and the kernel $k(\mathbf{x}, \mathbf{x'})$~\cite{Rasmussen2005-GaussianProcessesMachine}.
Given a set of previously evaluated points, the GP model predicts the value of the cost function over the entire parameter space and gives a measure of the model's uncertainty in less-explored areas. 
Key properties of the GP model, such as the smoothness of and variation between sampled functions are controlled via a set of hyperparameters, which are continually adjusted during training via maximum-likelihood estimation.
These hyperparameters include the noise level $\sigma_F$ of the data as well as the length scales $l$ which give the characteristic scale on which functions sampled from the GP change their value.
Control over these parameters can be a valuable tool to achieve optimum BO performance. For example, manually setting $\sigma_F = 0$ for training in the clean system can boost convergence and contribute to the low number of iterations needed compared to training in the noisy setting.
We also demonstrate how limiting the lower bound of $l$ -- constraining the maximum complexity of the constructed model -- can improve robustness while speeding up convergence in Appendix~\ref{subsec:UseModel}. \\
\\
After the surrogate model is initialized, a new point $\mathbf{x}_{i+1}$ is suggested based on the existing cost-function model, and the cost function is subsequently evaluated at this new point. The new tuple
\begin{align}
    (\mathbf{x}_i, C(\mathbf{x}_i))
\end{align}
is then used to update $\Tilde{C}(\mathbf{x})$, completing the loop. \\
The suggestion $\mathbf{x}_i$ is made by the acquisition function, which can prioritize either exploitation -- choosing an $\mathbf{x}_i$ which minimizes $\Tilde{C}(\mathbf{x})$, or exploration -- choosing an $\mathbf{x}_i$ where $\Tilde{C}(\mathbf{x})$ has high uncertainty.
In the lower confidence bound (LCB)~\cite{Srinivas2012-GaussianProcessOptimization} acquisition function
\begin{align}
    \mathrm{LCB}(x) = m(x) + \kappa \: \sigma_{\mathrm{GP}}(x)
\end{align}
used in this work, this tradeoff is tuned via the hyperparameter $\kappa$, which sets the weight of the model uncertainty $\sigma_\mathrm{GP}$.
In practice, we sweep between these two strategies, starting with a focus on exploration ($\kappa \sim 20$) and prioritizing exploitation ($\kappa \sim 0.2$) during the final iterations of the optimization. Fig.~\ref{fig_optimal_ramp}b) of the main text showcases $4$ sweeps of $\kappa$ and the corresponding evaluations of the cost function for a training run with a total of $500$ iterations. \\
To transfer results from a completed optimization run to a new one, e.g., to save cost when training on a noisy system, we replace random initialization with selected parameter-space points evaluated in the completed run. The points are selected by an algorithm detailed in a separate section. The corresponding values of the cost function are reevaluated in the new (e.g., noisy) environment, and the resulting (point, cost) tuples are used to initialize the surrogate model for the new optimization run.

\section{Exact Diagonalization Simulation}\label{subsec:ED}
Exact-diagonalization (ED) simulations are used to perform ground-state search and time evolution for the Harper-Hofstadter Hamiltonian~\cite{Harper1955-GeneralMotionConduction, Hofstadter1976-EnergyLevelsWave} on a $4 \times 4$ lattice with $N = 2$ bosons.
The parameters $t_{x / y}$ and $\Delta_{x / y}$ are the control fields tunable in the experiment -- representing tunneling amplitudes and global lattice tilts respectively. Site-dependent chemical potentials $\mu_{x, y}$ may be introduced to model disorder in the experiment. \\
For efficiency, we exploit the $\mathrm{U}(1)$ symmetry~\cite{Zhang2010-ExactDiagonalizationBosea}, resulting from particle number conservation.
As in the experiment, the initial state is prepared by manually placing the 2 bosonic particles in the bottom-left corner of the $4 \times 4$ lattice, i.e.,
\begin{align}
    \ket{\psi_\mathrm{initial}} = \hat{a}^\dagger_{0, 0} \hat{a}^\dagger_{0, 1} \ket{0},
\end{align}
where $\ket{0}$ is the $0$-particle state. \\
Time evolution is performed using the Runge-Kutta method RK5(4)~\cite{Dormand1980-FamilyEmbeddedRungeKutta} with a maximum step-size $\delta t_\mathrm{max} = 0.1$. \\
All computations for this work were run on a commercial laptop (Apple M1 Pro processor), where a single time evolution to $T = 100 \: \tau$ takes $\sim 3 \: \mathrm{s}$.
When running BO, fitting the surrogate model becomes more expensive than time-evolution simulation after about $200$ iterations. A typical optimization run to $500$ iterations takes about $1 \: \mathrm{h}$.

\section{Experimental Disorder Model}\label{subsec:Noise}
For a more realistic description of the experimental system, we consider disorder in the shape of static chemical-potential offsets that can be divided into two components: One as random offsets, and another as a harmonic confinement. \\
The random offsets arise due to light scattering on surfaces, introducing disorder in the optical lattice.
We model these offsets as uncorrelated with a maximum strength of $\sim 10-15 \: \mathrm{Hz}$, or about a quarter of the tunneling amplitude $\hbar / \tau \sim 40 \: \mathrm{Hz}$ in the final FQH state which sets our unit of energy.
The harmonic confinement is introduced by the optical potentials, or walls, that defines the region of interest. The walls are designed to have a Gaussian shape and are holographically projected by a digital micromirror device (DMD)~\cite{Zupancic2016-UltrapreciseHolographicBeam}. The DMD is located in the Fourier plane, and the finite aperture of the DMD limits the sharpness of the walls, resulting in edge offsets that may be modeled as a harmonic confinement.
\begin{figure}[t!!]
\centering
\includegraphics[width=\linewidth]{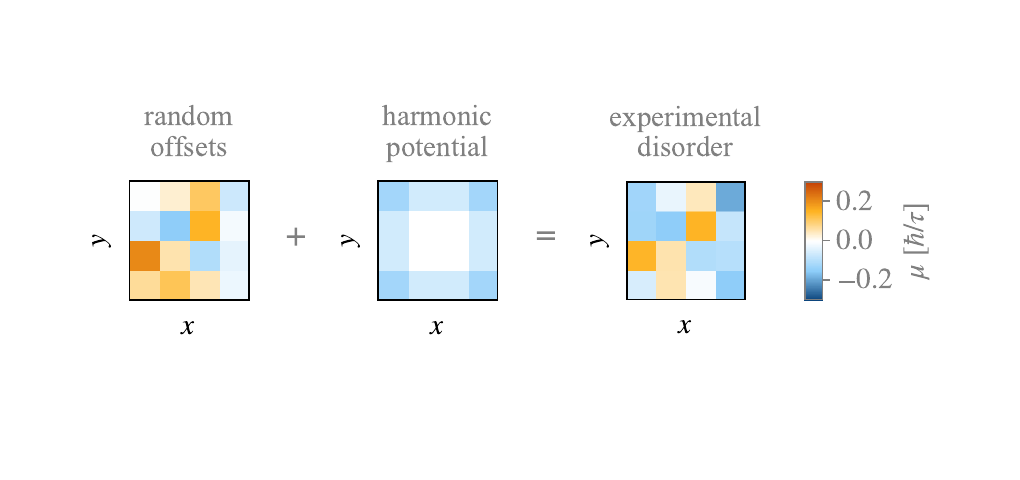}
\caption{{Experimental Disorder:}~Sample of a disorder realization with $\sigma = 0.1 \: \hbar / \tau$ composed of a random and a harmonic component. }
\label{supp_fig_experimental_noise}
\end{figure}
Thus, on the $4 \times 4$ lattice, the offset is modeled as having strength $2 \mu_\mathrm{DMD}$ in the corners of the lattice and $\mu_\mathrm{DMD}$ everywhere else on the edges.
The strength $\mu_\mathrm{DMD}$ of this type of disorder is taken to follow the same distribution as the random offsets, that is $\mu_\mathrm{DMD}$ may have either sign and the two types of disorder are similarly strong. \\
In the simulation, we model the disorder $\mu \ [\hbar / \tau]$ as being Gaussian-distributed with mean $0$ and consider a standard deviation of $\sigma = 0.1 \: \hbar / \tau$ to be the experimentally realistic disorder level.
For each numerical time evolution, we sample $4 \times 4 + 1$ independent values from this distribution -- one for each single-site offset and one for the global harmonic one. \\
We choose this disorder model as opposed to some other common experimental imperfections such as elevated temperatures and particle losses, as these can be well controlled in the small system by postselecting runs with the correct low-energy initial particle configuration and final particle number via density snapshots.

\section{Adaptive Sampling}\label{subsec:Sampling}
To keep the total number of samples used for training low, while reducing the noise of cost-function evaluations near the optima, we use a simple adaptive sampling technique -- increasing the number of samples based on the fidelity. Starting from a minimum number of $5$ samples, more samples are added if the averaged fidelity is above a threshold value.
The procedure continues with higher and higher fidelity thresholds until a predefined maximum number of $50$ samples.
We believe this technique to be beneficial as it allows us to more accurately map out the shape of discovered optima while saving cost by sacrificing accuracy in regions where the fidelity is low.

\section{Initialization Algorithm}\label{subsec:Algorithm}
To select the most valuable points in parameter space to transfer to a new optimization run, we use a simple heuristic algorithm. Iterating through all evaluated points, a point $\mathbf{x}$ is accepted, if it is more than a predefined distance $r$ away from all previously accepted points (in terms of the Euclidean distance in parameter space). If the closest accepted point $\mathbf{x}_\mathrm{acc.}$ is closer than $r$, between $\mathbf{x}_\mathrm{acc.}$ and $\mathbf{x}$, the point that yielded a higher fidelity is accepted.
The radius $r$ is chosen such that $\sim 100$ points are accepted in total. The $10 - 100$ points with the highest fidelity are then used to initialize the new optimization run. \\
This basic algorithm ensures that the point that yielded the highest fidelity in the BO run used for initialization is accepted and reevaluated for the new optimization. Apart from that, the algorithm prioritizes keeping points spread out over the parameter space (due to the division in `bubbles' of size $r$) while keeping points with high fidelity. We therefore expect this algorithm to succeed in selecting the most important local optima alongside the global optimum.
We expect it to work well, not only in our situation, where the location of the global optimum does not change much when introducing disorder to the system but also in more challenging situations, where the location changes significantly.
This is the case since, in general, we believe the global optimum in a disordered system to very likely have been a local optimum in the ideal system. We discuss this in more detail in the following sections.

\section{Brute-Force Search}\label{subsec:Brute-Force}
In the main text, using ramps with $3$ linear segments, we find the best performance in the disordered system with a ramp very similar to the optimum obtained in the clean system. To confirm that this is a feature of the problem setting and not a result of poor optimization performance, we perform a brute-force search in the simpler four-dimensional parameter space characterizing ramps with only $2$ linear segments. The fidelity with and without experimental disorder is performed within the search area
\begin{align}
    t_{x/y} \in [ 0.0, \, 1.2 ], \ \Delta_{x/y} \in [ 0.0, \, 4.0 ], 
\end{align}
stepping through a grid with step sizes $\delta t_{x/y} = \delta \Delta_{x/y} = 0.2$. We motivate this uniform grid size, taking more than $3$ times more points for the tilts $\Delta$ than for the hopping amplitudes $t$ by the increased sensitivity to tilts suggested by the GP models trained during BO (see main text).
Even if optimization performance were to be poor, we trust this general trend of sensitivities, as it is consistent in all BO training performed in this work.
To evaluate the fidelity in the presence of disorder, we use an adaptive sampling technique similar to the one used during noisy training -- ensuring high accuracy in high-fidelity regions while cutting down on the total number of evaluations.
Still, at least $10$ independent disorder realizations are evaluated for each point in the search grid, so more than $200 \, 000$ time-evolution simulations are carried out for the full brute-force search in the disordered setting.
\begin{figure}[t!!]
\centering
\includegraphics[width=\linewidth]{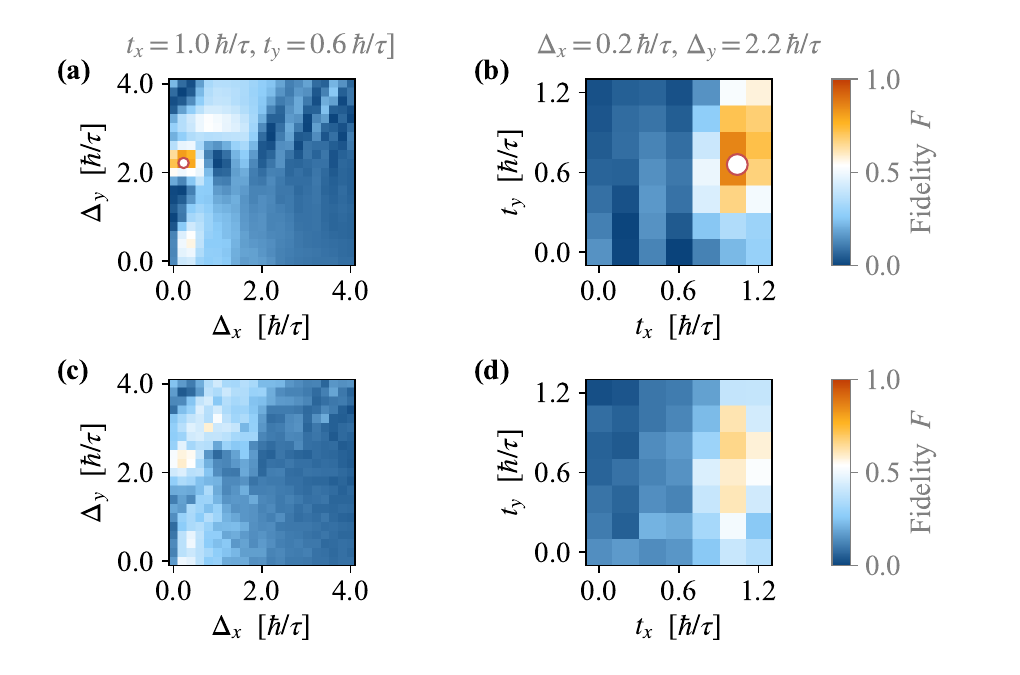}
\caption{{Brute-Force Search:}~The optimization landscape around the global optimum mapped out in the brute force search. \textbf{(a, b)} Ramp fidelity for variations of the tilts (a), and tunneling amplitudes (b) with the $2$ remaining parameters fixed to their optimum values (grey text).
The value of each pixel is obtained from a time-evolution simulation of the corresponding ramping protocol.
The white dots indicate the location of the optimum identified by BO.
\textbf{(c, d)} The same as (a, b), but for the brute-force search carried out in the disordered system with disorder strength $\sigma = 0.1 \: \hbar / \tau$. Each pixel represents the mean fidelity obtained from evaluating at least $10$ independent disorder realizations. }
\label{supp_fig_brute_force_landscape}
\end{figure}
Optimal fidelity is achieved at
\begin{align}
    [ \: t_x = 1.0, \ t_y = 0.6, \ \Delta_x = 0.2, \ \Delta_y = 2.2 \: ],
\end{align}
determined to be $F = 85.8 \%$ in the ideal system, and $F = 66(2) \%$ in the system including disorder.
Crucially, optimum fidelity with- and without disorder is reached at the same point in parameter space, so -- within the finite resolution provided by the grid -- the two optima coincide. \\
BO trained on the clean system with $500$ iterations finds the optimum to be at
\begin{align}
    [ \: t_x = 1.04, \ t_y = 0.66, \ \Delta_x = 0.23, \ \Delta_y = 2.21 \: ],
\end{align}
with $F = 87.8 \%$. These findings confirm that, in our problem setting, the global optimum of the cost landscape does indeed not move much with the introduction of disorder, and BO can efficiently and correctly identify said global optimum.

\section{Making Use of the Trained Model}\label{subsec:UseModel}
Here, we perform a closer investigation of how the surrogate model trained during a BO run can be used to make predictions about robustness and how training simpler models can lead to protocols that are more robust against unknown types of noise. \\
\begin{figure*}[t!!]
\centering
\includegraphics[width=\textwidth]{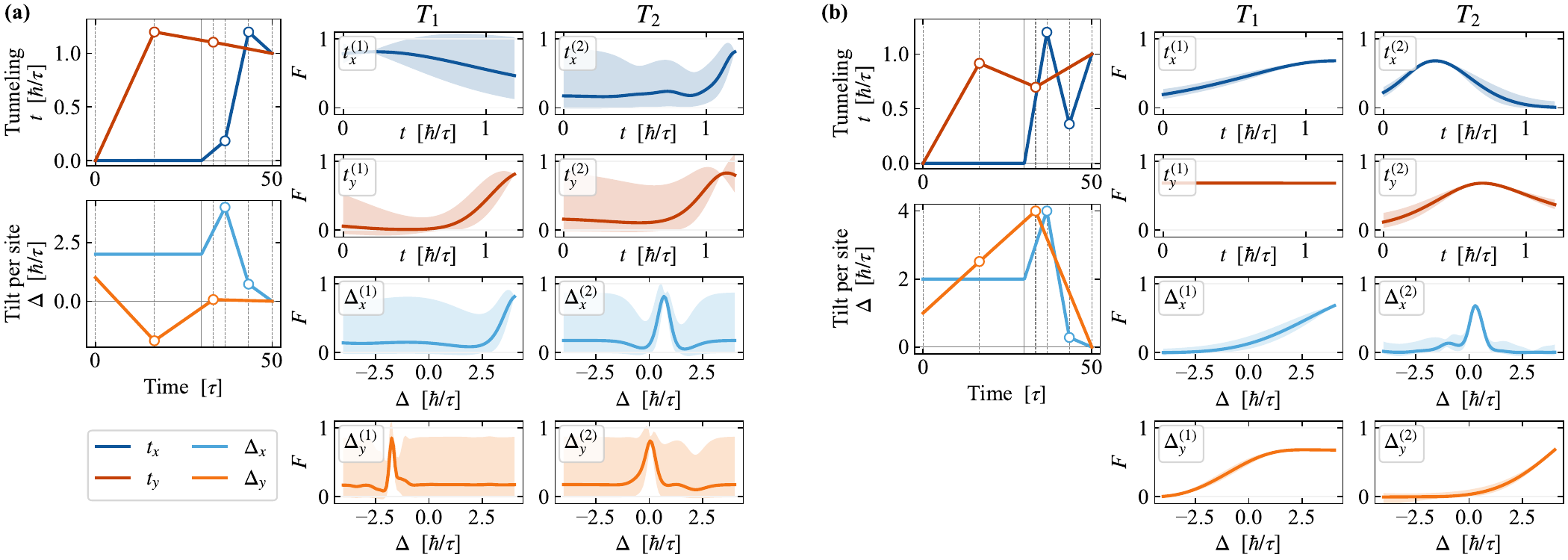}
\caption{{Different GP Models:}~Results of $2$ independent BO runs with $500$ iterations each for an intermediate ramp time $T = 50 \: \tau$. Based on different hyperparameter settings, the models identify different optima. Each subfigure ((a), and (b)) shows the optimized ramping protocol (left) and the modeled fidelity-landscape of the $8$ control parameters around the optimum (right).
\textbf{(a)} The model after training with the same parameter settings as in the main text. An optimum with fidelity $F = 81 \%$ is identified. The optimum is very narrow in parameter space and is especially sensitive to the $\Delta_y$ control points. The model uncertainties (shaded areas) away from the optimum are large.
\textbf{(b)} GP model with a $10 \times$ lower bound for the length scale hyperparameter than in (a). The lower-complexity model identifies a different optimum in the optimization landscape that lies lower in terms of fidelity ($F = 70 \%$) but is much wider. The model uncertainty around the optimum is very low.}
\label{supp_fig_fragile_gp_model}
\end{figure*}
Since the parameter space in the optimization directly translates to changes in Hamiltonian parameters over time -- just like many types of noise or disorder do -- we think it is reasonable to believe that the width of an optimum in parameter space can predict its performance against unknown types of noise. \\
If this relation between disorder in the experiment and noise in parameter space holds true, there is a simple intuitive picture of what is happening to the cost function when introducing noise to the system.
In common with other noisy signals, peaks in the optimization landscape would be expected to become wider and flatter with increasing disorder.
Depending on the type and strength of the noise, two distinct scenarios may occur:
\begin{enumerate}[label=(\roman*)]
    \item The global optimum widens but remains the highest peak, thus the location of the global optimum does not change much.
    \item Due to the loss in fidelity, a wider -- more robust -- local optimum becomes the new global optimum, thus the location of the global optimum in parameter space may jump dramatically.
\end{enumerate}
Based on the analysis presented in the main text, as well as the brute-force mapping of the cost landscape detailed above, we are confident that our system and disorder level realize scenario (i) -- both for long ($T = 100 \: \tau$) and short ($T \leq 20 \: \tau$) ramping protocols.
However, dramatic changes in the global optimum's location with the introduction of noise have been reported in the literature~\cite{Mukherjee2020-BayesianOptimalControl} indicating scenario (ii).
To look for (ii) in our system and test our conjecture, we turn to the intermediate ramp time $T = 50 \: \tau$. Here, in the clean system, BO identifies an optimum that it predicts to be extremely narrow in parameter space. The corresponding control path and GP model are illustrated in Fig.~\ref{supp_fig_fragile_gp_model}.
The figure also shows the model's uncertainty, which becomes high, even very close to the discovered optimum.
To accommodate the narrow optimum, the length-scale hyperparameters of the GPs become very low during training, making the uncertainty away from evaluated points high. \\
\begin{figure}[t!!]
\centering
\includegraphics[width=\linewidth]{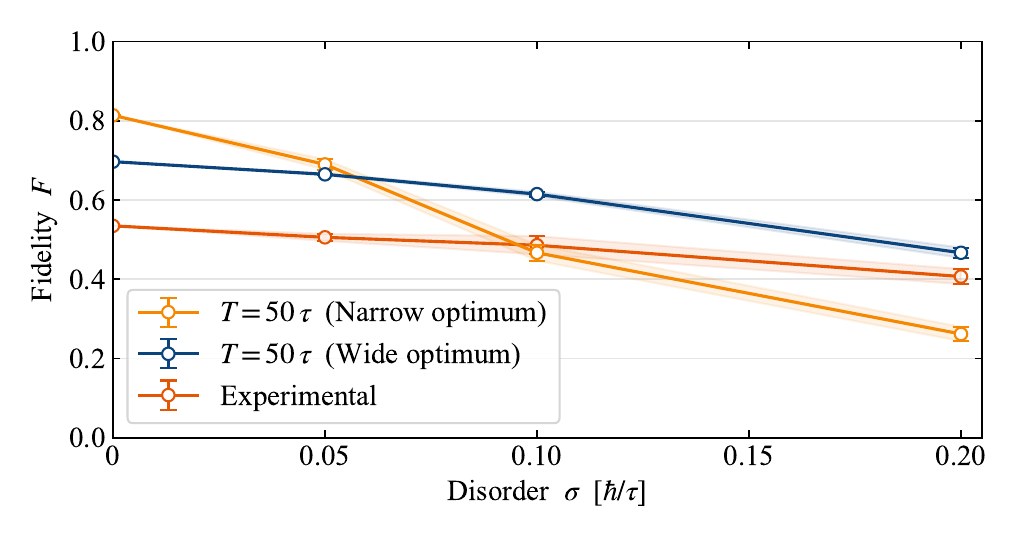}
\caption{{Robust Models:}~Fidelities of ramping protocols obtained from constrained and unconstrained training at varying disorder strength.
The error bars indicate the standard error of the mean fidelity obtained from averaging over $100$ disorder realizations.
While the ramp trained without any restrictions features the highest fidelity in the ideal system, it rapidly loses fidelity with the onset of experimental noise -- even falling below the manually designed protocol when crossing the experimentally realistic noise strength. The ramp trained with a simpler BO model with limited length scales is substantially more robust and retains its performance advantage over manual design in the presence of strong disorder. }
\label{supp_fig_robust_models}
\end{figure}
If there is indeed a direct relation between different types of noise, because the optimum is so narrow, we would expect this control path to be sensitive to unknown noises, such as the experimental potential noise used previously. This is confirmed in Fig.~\ref{supp_fig_robust_models}.
While the protocol evaluates to $F = 81 \%$ in the noiseless system, the fidelity quickly drops with the onset of noise and falls below that of the manually designed protocol for moderate to strong noise. \\
\\
We can go even further, and ask whether the GP model can not only predict poor performance against an unknown source of noise but if training can be actively shaped to avoid unsuitably narrow optima.
GPs provide natural approaches to tackle this issue, because they allow bounding the complexity of the model being constructed. In our example, by limiting the lower bound of the length scale hyperparameter, BO identifies an optimum that lies lower in terms of fidelity but is much wider in parameter space. \\
As hypothesized -- despite only knowing about the ideal system -- this new protocol is significantly more robust against the modeled experimental noise (see Fig.~\ref{supp_fig_robust_models}). It outperforms the more narrow optimum at experimentally realistic noise strengths in a meaningful way. The optimum strategy jumps from the narrow to the wider optimum between $\sigma = 0.05 \: \hbar / \tau$ and $\sigma = 0.1 \: \hbar / \tau$ which we interpret as a realization of (ii). \\
We conclude that the surrogate model has predictive power regarding an optimum's robustness to unknown sources of noise reiterating our recommended strategy to exploit the information contained in this costly model.
We also wish to highlight that limiting the model's complexity -- as has been done above using length scale bounds -- can not only produce more robust protocols but also improve the speed of convergence significantly, thus reducing the cost of training. This is apparent in the uncertainties of the models compared in Fig~\ref{supp_fig_fragile_gp_model} where the restricted model features substantially lower uncertainty after the same number of iterations. \\
\\
We conclude this discussion with some technical details about the occurrence of (ii) in our system (only) at intermediate ramp times. We attribute this curious behavior to a qualitative transition in the optimal ramping protocol causing issues due to the simple parametrization using linear segments between fixed points in time. In the main text -- for both the experimental and the optimized protocol -- we observed a crossing in the control fields $t_x(t)$ and $t_y(t)$, where $t_x$ is ramped up from $t_x = 0.0$ to more than unity and subsequently decreased back to $t_x = 1.0 \: \hbar / \tau$. This feature is absent in the $T = 20 \: \tau$ ramp -- presumably due to a lack of time. Consequently, there must be a crossover at an intermediate total time $20 \: \tau \leq T \leq 100 \: \tau$ where both strategies are equally good. This complicates training and is likely exacerbated by the rigid parametrization which does not allow `doing nothing'. Similar to longer protocols, the optimized result for $T = 50 \: \tau$ shown in Fig.~\ref{supp_fig_fragile_gp_model} features a crossing $t_x(t) = t_y(t)$ at $0 < t < T$ but has to speed up delocalization in the y-direction by ramping $\Delta_y$ to a value significantly below $0.0$. Consequently, the control path becomes extremely sensitive to this $\Delta_y$ value which turns out to be the optimization parameter that is assigned the shortest length scale. \\

\section{Larger System}\label{subsec:Pfaffian}
\begin{figure}[t!!]
\centering
\includegraphics[width=\linewidth]{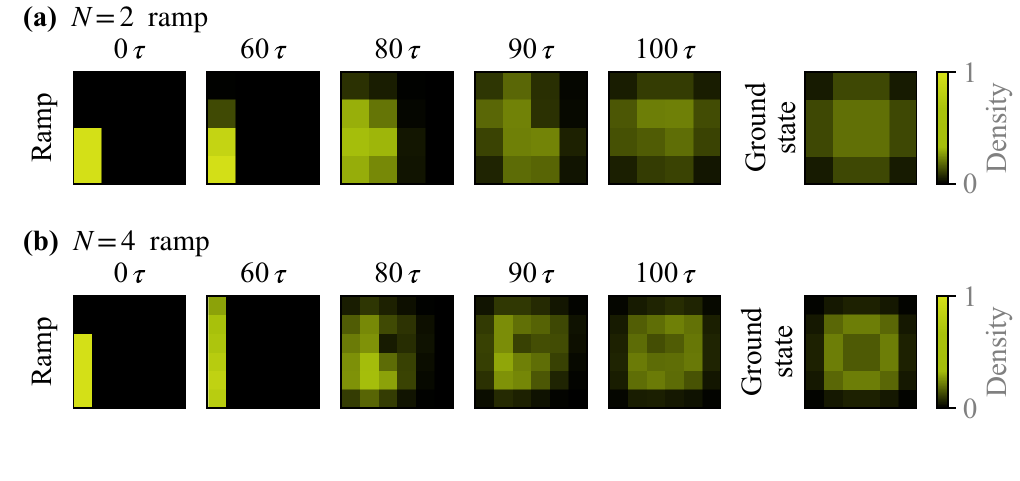}
\caption{{Particle Density:}~Evolution of the density profile from the initial state of pinned particles comparing different system sizes. The plots on the left show the density distribution at different times during the ramp. The rightmost plot shows the density profile of the (target) ground state. \textbf{(a)}~Smaller system of $2$ particles on $4 \times 4$ sites evolving according to the optimized protocol from Section~\ref{sec:Optimization}.
\textbf{(b)}~Larger system of $4$ particles on $6 \times 6$ sites evolving according to the optimized protocol obtained via transfer learning. }
\label{supp_fig_optimal_ramp_pfaffian}
\end{figure}

As an outlook beyond the minimal two-particle Laughlin-like state at filling factor $\nu = 1/2$ that has been realized experimentally, we conclude with first results on the $N = 4$, $L_x = L_y = 6$ Pfaffian-like state at $\nu = 1$ proposed by F. Palm et al.~\cite{Palm2021-BosonicPfaffianState}.
Following their work, we use the system parameters $U = 4 \: \hbar / \tau$ and $\phi / 2 \pi = 0.16$.
We then optimize the ground-state preparation from an initial state of particles pinned in the corner of the system in a $1 \times 4$ configuration.
\begin{figure}[t!!]
\centering
\includegraphics[width=\linewidth]{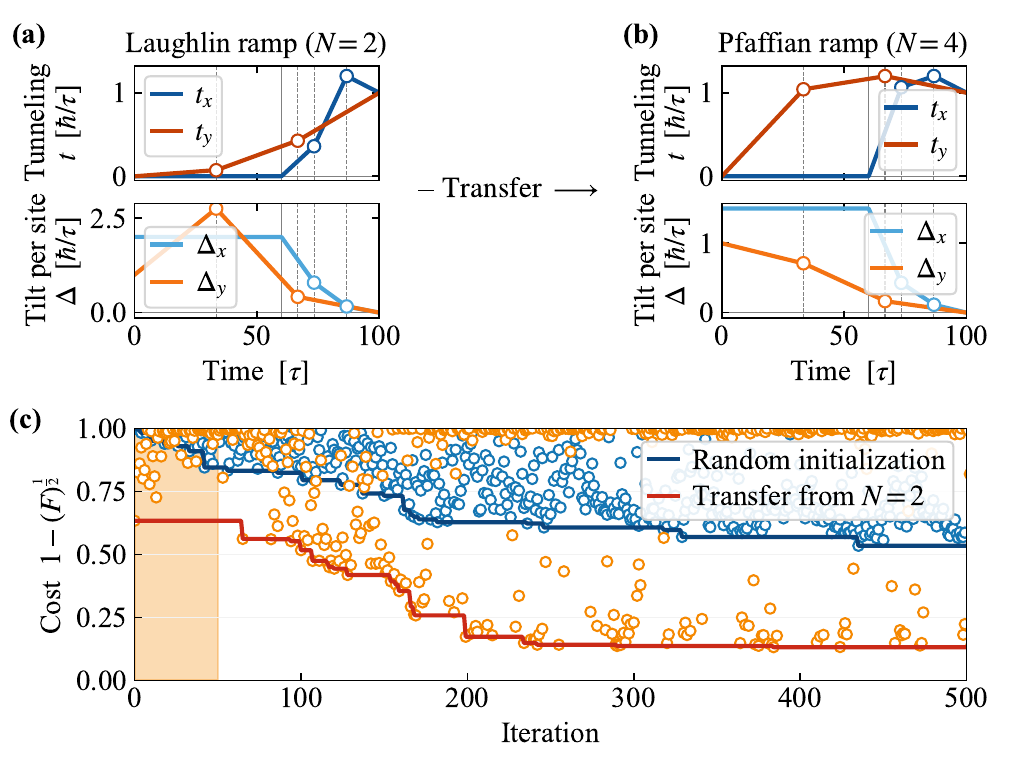}
\caption{{Transfer to Larger System:}~BO training for the $N = 4$, $L_x = L_y = 6$, $\nu = 1$ state. \textbf{(a)}~The optimized protocol for the two-particle $\nu = 1/2$ state from Section~\ref{sec:Optimization}. \textbf{(b)}~Optimized ramping protocol for the larger $\nu = 1$ state obtained by transferring training data: $50$ protocols from the training of (a) are used to initialize the surrogate model. \textbf{(c)}~Comparison between training initialized with $50$ transferred points and random initialization with $10$ points. The transfer approach is much more efficient, reaching $F > 70 \%$ in less than $200$ iterations. The shaded orange region marks the evaluation of transferred data points. The optimized Laughlin ramp prepares the larger $\nu = 1$ state with $F=13 \%$ ($\mathrm{Iteration} = 0$; orange) -- significantly better than random protocols, but far from optimal. }
\label{supp_fig_pfaffian_transfer}
\end{figure}
While pinning the particles in a corner of the system becomes increasingly restrictive when going to larger systems, we choose this state to most naturally extend the optimization problem from the $N = 2$ case. 
Combined with the larger Hilbert-space dimension $\mathrm{dim}(\mathcal{H}^{N=4}) = 82251$, compared to $\mathrm{dim}(\mathcal{H}^{N=2}) = 136$, this makes for a significantly more challenging preparation task, where randomly chosen protocols are expected to yield almost vanishing fidelity.
As the goal of this work is to allow the preparation of more complex states using contemporary experimental setups, we perform optimization with the same total protocol duration and restrictions imposed by Floquet heating as previously for the smaller system.
That is, we again optimize protocols of total duration $T = 100 \: \tau$ with the coupling in x-direction only active for a shorter time $T_x = 40 \: \tau$.
We wish to answer two main questions:
\begin{enumerate}[label=(\roman*)]
    \item Can we still find good preparation protocols for the larger system? How do these compare to the adiabatic protocol for the smaller system?
    \item Can we use the knowledge gained in the small system to boost optimization efficiency for the more complex state?
\end{enumerate}
In order to answer these questions, we perform two separate BO runs with $500$ iterations each.
The first run follows the procedure described in Section~\ref{sec:Optimization}, using $10$ random initial points, while the second run uses initial points collected from training on the small systems following the initialization algorithm described in Section~\ref{subsec:Algorithm}.
The performance of these two approaches is compared in Fig.~\ref{supp_fig_pfaffian_transfer}.
While the `cold-start' approach that uses random initialization continues to find better and better protocols as training goes on, the rate of improvement is low, reaching only $F = 22 \%$ after $500$ iterations.
The transfer approach performs much better, reaching fidelities as high as $F = 75 \%$.
While the optimized protocol for the $N=2$ system is not directly applicable to the larger system ($F = 13 \%)$, the qualitative strategy does not change. Fig.~\ref{supp_fig_optimal_ramp_pfaffian} compares the evolution of the density profile for the optimized protocol in the two system sizes. \\
The optimization performance was boosted massively by utilizing knowledge of the smaller system. This extends a previous report of efficient scaling in the system size~\cite{Xie2022-BayesianLearningOptimal} to a case featuring different model parameters ($U$ and $\alpha$) and a different target state.
In summary, the transfer technique allows us to positively answer both (i) and (ii).
For our system, optimization allows preparing a significantly more complex state at a higher fidelity than the manual benchmark on the small system with the same experimental constraints.
For reference, applying the manual protocol to the $N=4$ system results in $F = 3.4 \%$ while a simple linear ramp for all control fields has $F = \num{9.7e-7}$.
Additionally, optimization on a small system easily tractable using numerics was able to dramatically reduce the costs of optimization on a much more demanding system.

\bibliographystyle{quantum}
\bibliography{optimization_paper}

\end{document}